# Pressure-mediated crystalline g-$C_3N_4$ with enhanced spatial charge transport for solar $H_2$ evolution and photocathodic protection of 304 stainless steels

Xiaochun Gao, Ludong University, Laboratory of Plasma and Energy Conversion, School of Physics and Optoelectronic Engineering, Ludong University, Yantai, China.

Shaoqi Hou, School of Mathematical and Physical Sciences, Faculty of Science, University of Technology Sydney (UTS), Sydney, NSW, 2007, Australia.

Dawei Su, School of Science, STEM College, RMIT University.3001 Melbourne, Australia

## 1. Introduction

Since the groundbreaking work in 2009,[1] the conjugated polymerized g-$C_3N_4$ has become one of the most promising semiconductors in converting solar to other types of renewable energy owing to the merits of unique electronic band structure, stable physicochemical properties, and environmental benignity.[2–7] Researchers have emphasized the important role of defect engineering in ameliorating the insufficient solar absorption and sluggish photocarrier transfer kinetics of g-$C_3N_4$ in terms of tailoring its band structure, charge distribution as well as electronic conductivity.[8–11] However, the defects might cause excessive dangling bonds and lattice disorder, which would deteriorate the surface states and destroy the conjugated system of g-$C_3N_4$, leading to severe photocarrier recombination and reduced light-harvesting ability.[8] Thus, the defects should be properly modulated on the surface of g-$C_3N_4$ to avoid the dramatically reduced amount of thermodynamically active photogenerated electrons.

High crystallinity, as the complementary part to defects, has also been demonstrated to be efficient in boosting the photocatalytic activity of g-$C_3N_4$.[12–16] Compared to the traditional thermal polymerized g-$C_3N_4$ with low crystallinity, the high crystalline g-$C_3N_4$ exhibits superior performance in spatial charge transport.[17,18] Specifically, for the in-plane photocarrier transport, the infinitely extended tri-s-triazine-based or triazine-based high crystalline g-$C_3N_4$ could stabilize the π-electron conjugated system, which guarantees fast charge mobility along the one-dimension direction.[19] Additionally, prior theoretical and experimental studies have also evidenced a higher in-plane crystallinity could lower the band gap of g-$C_3N_4$, further promoting its solar harvesting capability.[18] Notably, accelerated interlayered charge transport in high crystalline g-$C_3N_4$ can also be achieved as there are fewer melon oligomers and H bonds (N ⋯ H), which can avoid the high localization of photocarriers at the terminal sites and substantially alleviate the blocking of electron conduction across the plane.[14] Therefore, the crystallinity regulation of g-$C_3N_4$ is particularly of vital importance for spatial charge transfer and surface redox reaction, which should be paid more research attention to.

Generally speaking, the crystalline g-$C_3N_4$ can be successfully obtained via the ion thermal reaction in the presence of melting salts (LiCl, NaCl, KCl) as their solid surfaces have a critical role in accelerating the kinetics of precursor diffusion, polymerization, and preferred facets orientation.[14,16,20–23] For instance, under vacuum thermal condensation, the KCl/LiCl eutectic mixture was employed to fabricate the poly (triazine imide) (PTI)-based g-$C_3N_4$ single crystals with preferred prismatic {10$\bar{1}$0} planes, which were demonstrated to be the major reactive facet toward overall water splitting.[16] In contrast, under a normal pressure environment, the KCl/LiCl promises the formation of poly (tri-s-triazine imide) (PHI)-based g-$C_3N_4$, which is superior to PTI-based g-$C_3N_4$ toward various solar activities.[15,24] In a following-up study, the in-plane highly ordered g-$C_3N_4$ with sharp (100) facet was obtained using NaCl as the sole salt,[19] giving rising to the fast charge mobility and high photocarrier separation. Intriguingly, these alkali metallic ions ($Li^+$, $Na^+$, $K^+$) also inevitably induce structural defects such as cyano groups (-C≡N) and limit the amount of amino groups (-$NH_x$, x=1, 2) during the ion thermal polymerization. This might provide us with an opportunity to develop functional crystalline g-$C_3N_4$ while simultaneously maintaining the optimized defects via a modified ion thermal reaction, which has been less emphasized in previous reports.

Herein, we proposed a pressure-mediated ion thermal synthetic strategy to obtain the high crystalline g-$C_3N_4$ with optimized surface defects (**Figure 1**). The NaCl/KCl eutectic mixture was chosen as the pyrolysis solvent to enhance the crystallinity along both the in-plane and cross-plane directions, which is believed to facilitate charge transport under irradiation. While the high pressure inside the precursor tablet further induces a narrower interlayered distance, shortening the pathways of both diffused active species and the excited photocarriers, and thus renders g-$C_3N_4$ with accelerated redox kinetics. More importantly, the pressure-medicated crystalline g-$C_3N_4$ (CCN-P) also has been verified to maintain the surface defects (mainly referring to -C≡N and -$NH_x$) in a more reasonable level in comparison to the crystalline g-$C_3N_4$ without high-pressure regulation (CCN-NP). As a result, CCN-P exhibits the highest electron-trapping resistance ($R_{trap}$) up to 11.36 kΩ $cm^2$ and the slowest photocarrier decay kinetics of 0.013 $s^{-1}$, which are 3.5 times higher and 14.6 times lower than those of its counterparts. Along with its enhanced light harvesting ability, CCN-P delivers a superior photocatalytic $H_2$ evolution rate of 2168.8 μmol $g^{-1}$ $h^{-1}$ and excellent photocathodic protection of 304 stainless steel (304 ss) with the highest dark protection efficiency of 78.5% after 7500 s, far more exceeding the bulk g-$C_3N_4$ and CCN-P.

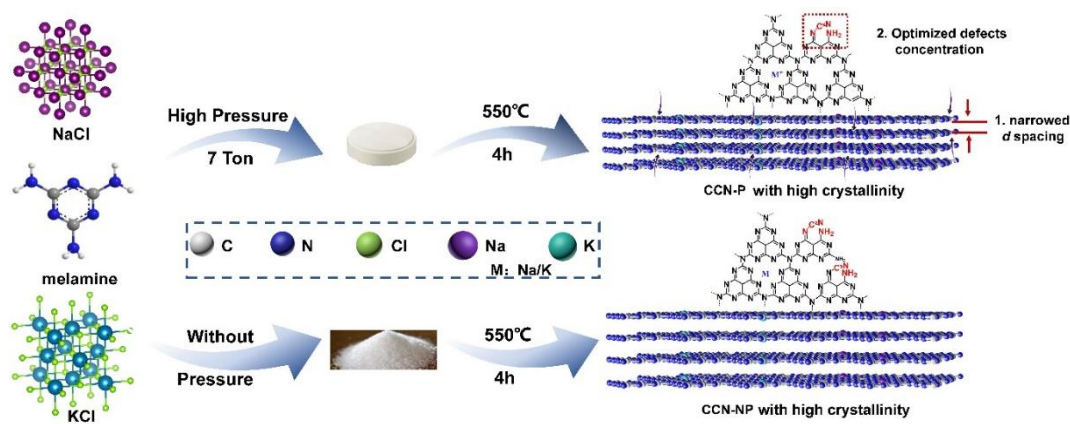

**Figure 1** Schematic illustration for the synthesis of high crystalline CCN-P and CCN-NP with/without high pressure.

## 2. Results and Discussion
### 2.1 Morphology and Structure

Inspired by the crystalline g-$C_3N_4$ synthesized via the ion thermal reaction under vacuum conditions or normal pressure,[15,16,24] we proposed a facile synthetic strategy to evaluate how the crystalline orientation would change under high pressure. As shown in **Figure 1**, melamine and NaCl/KCl eutectic mixture were firstly well mixed and then pressed into a white tablet with a diameter of 12 mm and a height of 5 mm under a pressure of 7 tons for 5 min. After a thermal treatment at 550 °C for 4 h and a subsequent washing process, the resultant yellowish CCN-P powder was collected. Considering the photocatalytic activity of crystalline g-$C_3N_4$ obtained via a vacuum condition is relatively poor or ignorable, and thus we designed the controlled samples, which were synthesized under normal pressure in the presence of NaCl/KCl salts (CCN-NP) or via a traditional annealing method with sole melamine (CN-B), respectively.

The crystallinity of these g-$C_3N_4$ can be clearly reflected by the high-resolution TEM (HRTEM) images. As shown in **Figure 2a**, CN-B is composed of irregular aggregates with very thick laminar layers. Additionally, no obvious lattice fringe is observed (**Figure 2b**), indicating the traditional thermal condensation only leads to the amorphous g-$C_3N_4$ without lattice order, which is consistent with previous reports.[25,26] Unlike other fabricated crystalline g-$C_3N_4$ using pristine g-$C_3N_4$ as the precursor,[13–15] our synthesized CCN-NP and CCN-P employing melamine as precursor also display a high crystallinity. In detail, CCN-NP exhibits a compact and assembled nanorod structure with a subunit size of ~200 nm (**Figure 2c**). These obvious lattice fringes with a *d* spacing of 1.197 nm are assigned to the (100) planes, which confirms the presence of NaCl and KCl indeed enhanced the in-plane polymerization for CCN-NP (**Figure 2d**). Interestingly, CCN-P displays a nanoplate-like structure in micrometre size (**Figure 2e**), which suggests the high pressure might favour the crystals' growth along the in-plane direction. Notably, in comparison with CCN-NP, CCN-P reveals a narrower $d_{(100)}$ spacing of 0.939 nm, confirming the high strain inside the precursor is efficient in shortening the in-plane distance of repeated units, which is inferred to accelerate the internal charge transfer.

In good accordance with TEM results, the XRD patterns also evidence the higher crystallinity of CCN-P and CCN-NP. An obvious and strong (100) crystal plane at around 8.0° is observed for both CCN-P and CCN-NP, confirming the repeated in-plane tri-s-triazine unit can be well preserved under high pressure. This confirms the NaCl/KCl mixture is critical for the enhancement of in-plane polymerization, suggesting an extended conjugated π electronic system that would boost the charge transfer in one dimension. While for CN-B, the broad and weakened (100) plane is presented with a red shift by 5.09°, which is supposed to be its low polymerization degree and low in-plane lattice order due to the edged amino groups. Importantly, the (002) plane of CCN-P shifts to a much higher angle by 0.6° compared to the (100) plane of CCN-NP (**Figure 2h**), suggesting the high pressure is more favourable in narrowing the interlayer distance than the in-plane distance. Thus, for CCN-P, the photocarrier transport pathway along the cross-plane direction can be significantly shortened, and higher solar activity is also expected.

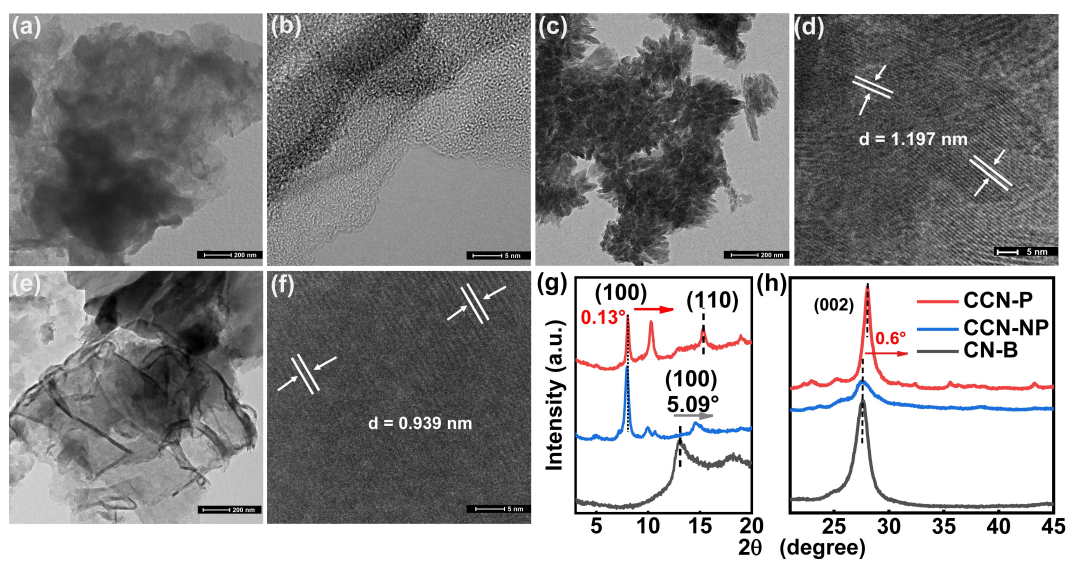

**Figure 2** TEM images of **(a, b)** CN-B; **(c, d)** CCN-NP; **(e, f)** CCN-P; XRD patterns of g-$C_3N_4$ samples with the 2θ range of; **(g)** 3-20°; and **(h)** 20-45°.

**2.2 FT-IR, EPR, and XPS analysis**

The surface functional groups of the as-prepared CCN-P, CCN-NP, and CN-B were investigated by the Fourier transform infrared technique (FT-IR). As seen from **Figure 3a**, both CCN-P and CCN-NP show the absorption bands at 807 cm$^{-1}$ and 1240-1621 cm$^{-1}$, corresponding to the breathing mode of tri-s-triazine and vibration mode of C=N-C absorption bands, respectively.[26] This observation indicates that the ion thermal reaction performed at normal and high pressure could maintain the basic framework of heptazine-based g-$C_3N_4$, showing no difference to CN-B obtained via the traditional thermal condensation method. However, CCN-P and CCN-NP exhibit a new absorption band at around 2177 cm$^{-1}$, which is assigned to the cyano groups (-C≡N). This is in good line with previous reports that the addition of alkali-containing compounds (such as NaOH, KOH, Ba(OH)$_2$, NaCl, KCl, LiCl)[7,27–30] into the thermal polymerization of melamine-like precursors would always induce the formation of -C≡N groups, probably via the Mars van Krevelen (MvK) mechanism.[31,32] It is worth mentioning that -C≡N intensity of CCN-NP is much higher than that of CCN-P, implying the precursor in a loosened powder without strain might favour the formation of surface functional groups. In addition, the lowest peak intensity at around 3000-3400 cm$^{-1}$ of CCN-P also demonstrates the lowest amount of -NH$_x$ or absorbed O-H groups,[8] which further evidence the pressure-mediated ion thermal reaction could tailor the defective surface groups into a reasonably lower level, providing the research space for defect controls on g-$C_3N_4$.

The electron paramagnetic resonance (EPR) is employed to trace the vacancies formation of the as-prepared g-$C_3N_4$ samples. As shown in **Figure 3b**, all photocatalysts display one single Lorentzian line with a g value at around 2.0037, which is ascribed to the lone pair electrons of sp$^2$ hybridized C in the aromatic rings.[8,11,33] It is reasonable to observe that CCN-P has a slightly lower EPR signal than CN-B, which is attributed to its intrinsic lower defect density and the counteraction effect on sp$^2$ hybridized C concentration induced by the small amount of sp hybridized -C≡N groups.[32] Yet, owing to the highest -C≡N concentration, CCN-NP exhibits an inverse and highest EPR signal, which is supposed to be the formation of N vacancies. This is also in good accordance with the FT-IR result that the high-pressure regulated ion thermal pyrolysis enables a lower defect density compared to the counterpart in normal pressure.

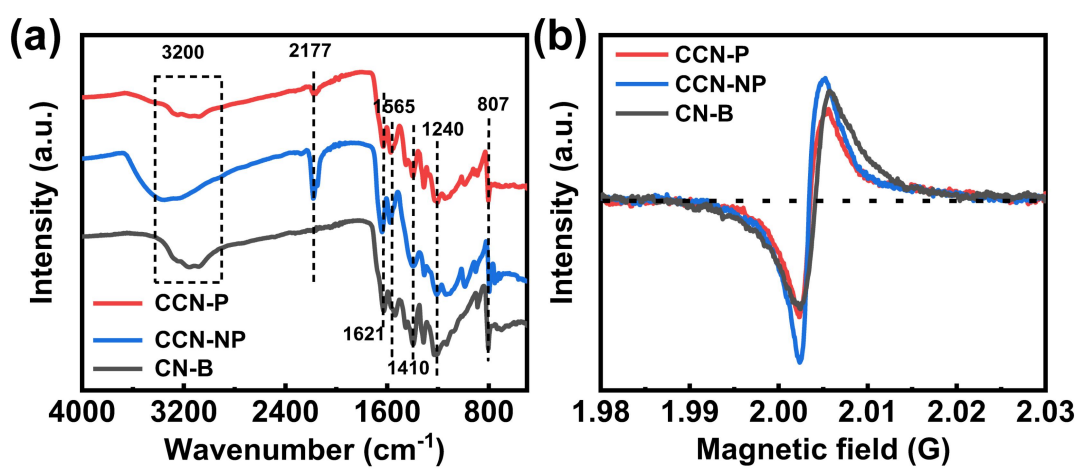

**Figure 3 (a)** FT-IR spectra and **(b)** EPR spectra of CCN-P, CCN-NP, and CN-B.

X-ray photoelectron spectroscopy (XPS) was used to probe the chemical state of elements change under high-pressure regulation. The survey scan spectra in **Figure 4a, b** clearly shows the signal of Na 1s and K 2p for CCN-P and CCN-NP after ion thermal reaction, which is consistent with previous literature.[34–37] Furthermore, the corresponding Na and K doping levels are similar at around 1.66% vs. 1.47% and 1.78% vs. 1.77% (**Table S1**), indicating that high pressure has an ignorable effect in alkali ion doping. As for the C 1s spectrum (**Figure 4c**), all samples exhibit the typical C-C (C1) and C=N-C (C3) peaks at around 284.5 and 287 eV, respectively (**Table S2**).[8,26,38] In addition, an obvious XPS peak shift to lower binding energy by 0.8 eV of crystalline g-$C_3N_4$ is also observed, which is supposed to be the electron-donating effect from K$^+$ or Na$^+$ to C atoms within the significantly enhanced polymerized tri-s-triazine rings. Interestingly, a new deconvoluted peak at 285.4 eV, assigning to the -C≡N species (C2),[30,32] is founded for the crystalline g-$C_3N_4$. Consisting with the FT-IR result, CCN-P displays a lower-C≡N concentration of 8.9% than CCN-NP, revealing the critical role of high strain within the precursors in suppressing the defects formation.

As for the N 1s spectrum (**Figure 4d**), four main peaks at around 398.0, 399.0, 400.0, and 403.5 eV were divided, which originate from the sp$^2$ hybridized C-N=C (N1), sp$^3$ hybridized $C_3$-N (N2), amino groups carrying hydrogen C-N-H (N3), and π excitation (N4), respectively.[39] Similar to the C 1s spectra shifting, the N species in

CCN-P and CCN-NP also reveal a blue shift owing to the electron donation of alkali metal induced by electron discrepancy. However, compared to CN-B, one can see a slight N1 blue shift by 0.1 eV for CCN-NP, while this value is enlarged to 0.4 eV for CCN-P. The reason is ascribed to the stronger electron-withdrawing effect of more -C≡N groups in CCN-NP than that in CCN-P.

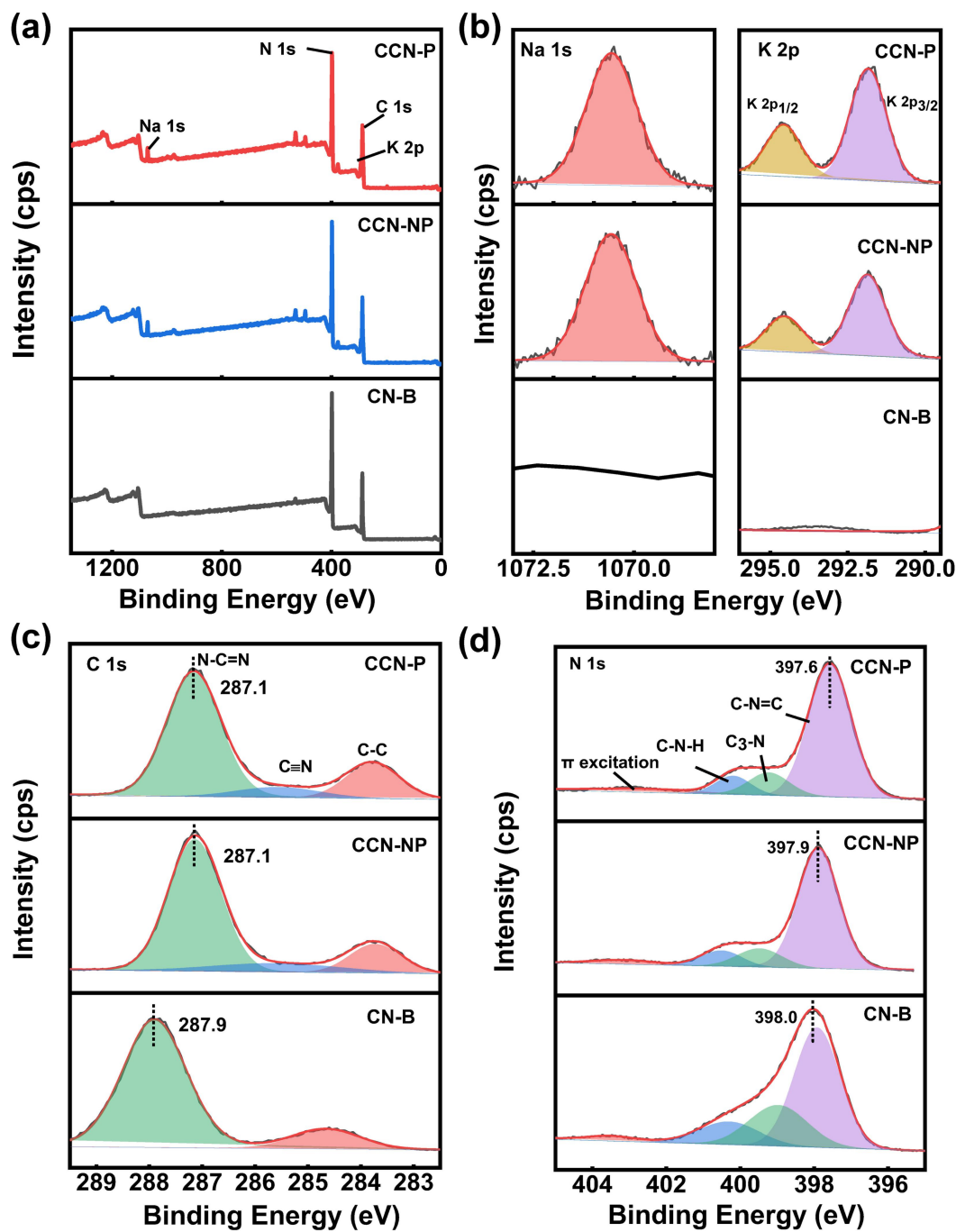

**Figure 4** **(a)** Survey XPS spectra; Core-level XPS spectra of **(b)** Na 1s, K 2p; **(c)** C 1s and **(d)** N 1s for CCN-P, CCN-NP, and CN-B.

## 2.3 PL, DRS, and Electronic Band analysis

The time-resolved PL spectra were recorded to reveal the photocarrier transfer process of CN-B, CCN-NP, and CN-P, respectively. Generally speaking, a shorter lifetime represents a fast photocarrier separation and transfer rate, thus reducing the chance to be recombined either by the radiative fluorescence or the non-radiative heat. As shown in **Figure 5a**, all samples show a non-exponential decay curve which can be fitted by the decay function of three exponents: [40,41]

$$I(t) = B + A_1 \exp(-t/\tau_1) + A_2 \exp(-t/\tau_2) + A_3 \exp(-t/\tau_3)$$

Where $I$ is the fluorescence intensity, B and $A_1$, $A_2$, and $A_3$ are constants that were obtained after fitting every decay curve. The fast ($\tau_1$) and slow ($\tau_2$, $\tau_3$) decay components were then generated. The former is associated with the photocarrier recombination process, and the latter ones are associated with non-irradiative recombination. Additionally, the average lifetime ($\tau_{ave}$) is usually used to evaluate the overall lifetime of samples under irradiation, which can be calculated by following equation: [40,41]

$$\tau_{ave} = \frac{A_1\tau_1^2 + A_2\tau_2^2 + A_3\tau_3^2}{A_1\tau_1 + A_2\tau_2 + A_3\tau_3}$$

According to the fitting constants in **Table S3**, the CNN-P exhibits the shortest $\tau_{ave}$ of 0.61 ns, which is 14.39 and 1.80 times shorted than CCN-NP and CCN-P, further indicating the favorable fast charger transfer progress induced by the pressure regulation (**Figure 5a**). The UV-vis DRS also confirms the promoted visible light absorption of crystalline g-$C_3N_4$. Specifically, in increasing order, CN-B, CCN-NP, and CCN-P exhibit extended visible adsorption edges of 464, 469, and 481 nm, respectively (**Figure 5b**). Employing the Kubelka-Munk method,[8] we obtained the smallest bandgap of 2.67 eV for CCN-P, which is 0.05 and 0.13 eV lower than those of CCN-NP and CN-B (**Figure 5c**), further indicating the efficiency of our pressure-mediated strategy toward light harvesting.

To describe the band alignments, we determined the CB positions of CN-B, CCN-NP, and CCN-P by Mott-Schottky measurement (**Figure 5d**). The values are -1.23, -1.13, and -1.21 V *vs.* AgCl/Ag, which can be converted to -0.62, -0.52, and -0.60 V *vs.* reference hydrogen electrode (RHE), respectively. Combining the bandgap values (**Figure 5c**), we draw the electronic band structures of the various g-$C_3N_4$ samples in **Figure 5e**. Notably, one can see there are almost no prominent defect states after the ion thermal reaction for both CCN-NP and CCN-P despite the new -C≡N groups appearance, which might be the overall lower defect concentrations in comparison with those for other defect controlling strategies in the absence of ionic salts. In summary, the ion thermal synthetic approach regulated by pressure could not only extend the light-harvesting ability but enhance the photocarrier separation efficiency of crystalline g-$C_3N_4$, which is inferred to boost the following photoelectrochemical performance and solar hydrogen evolution.

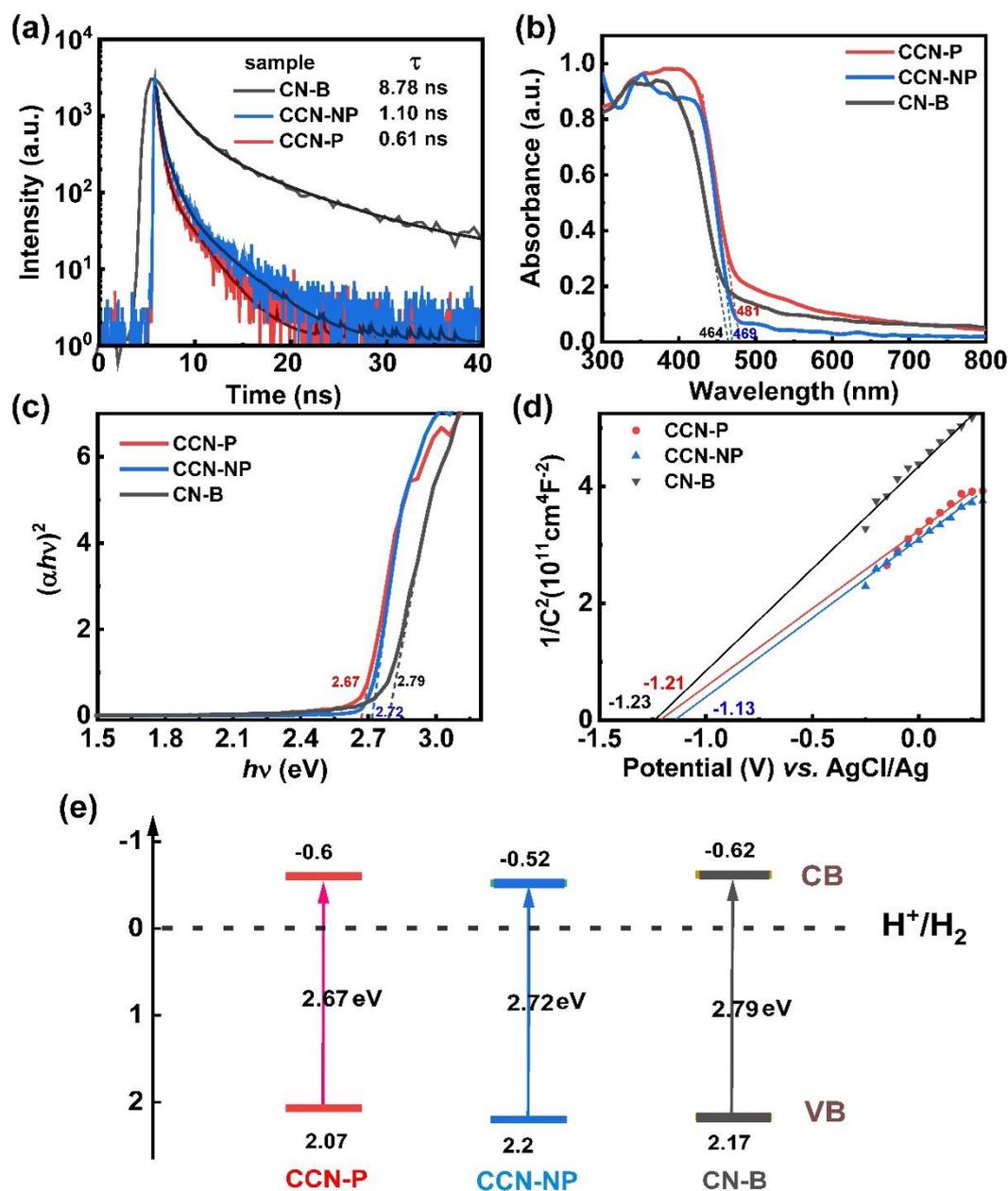

**Figure 5 (a)** Time-resolved fluorescence spectra; **(b)** UV-visible diffusion reflectance adsorption spectra; **(c)** (αhν)$^2$ *vs.* hν plots based on Kubelk-Munk method; **(d)** Mott-Schottky curves and **(e)** Band structures of CCN-P, CCN-NP, and CN-B.

**2.4 Photocatalytic performance**

Different from the above qualitative PL analysis, the quantitative photoelectrochemical (PEC) measurements probing the photocarrier transport were performed in a typical three-electrode cell system with the back-illuminated g-C$_3$N$_4$-based photoelectrodes as working electrodes under a chopped simulated solar light. Considering the distinct electron behaviors in dark and under irradiation, we employed different equivalent circuits to interpret each charge transfer progress according to the EIS spectra (inset, **Figure 6a-b**), and the fitting parameters are shown in **Table S4** and **Table S5**. As the first step of excited photocarrier transport, the bulk charge transfer resistance ($R_{ct,bulk}$) is used to reflect the charge transfer barrier from bulk to the surface. We observed that CCN-P, CCN-NP, and CN-B all exhibited a dramatically reduced bulk charge transfer resistance ($R_{ct,bulk}$) value of $4.63 \times 10^3$, $6.56 \times 10^3$, and $2.55 \times 10^4$ Ω cm$^2$ under solar irradiation than those in dark, implying their enhanced PEC activity. Among these, CCN-P delivered the largest $R_{ct, bulk}$ decreasing rate up to 92.6%, which is beneficial to the fastest photocarrier mobility from bulk phase to the surface. Based on its lowest bulk charge transfer barrier and most suppressed photocarrier recombination (**Figure 5a**), we speculate that CCN-P has the best bulk charge transport path, which might be ascribed to the narrower interlayered space induced by the high-pressure regulation.

The last step of photocarrier transport from the surface of CCN-P, CCN-NP, and CN-B to the electrolyte is much more complex as the electrons are vulnerable to recombine with holes due to the abundant surface dangling bonds, including the -NH$_x$ and -C≡N groups. To further compare the ease of electrons escaping from these dangling bonds, the charge trapping resistance ($R_{trap}$) was introduced to the equivalent circuit,[42] which a lower value indicating the deteriorated surface states with electrons prone to being deeply trapped and recombining with holes. As shown in **Figure 6e** and **Table S5,** CCN-P exhibits the highest $R_{trap}$ of $1.14 \times 10^4$ Ω cm$^2$, which is 2.0 and 3.5 times higher than CCN-NP and CN-B, further demonstrating its superior interfacial charge transfer. This might be ascribed to the high-pressure tailored high crystalline CCN-P with appropriate surface defect levels of both -C≡N and -NH$_x$ groups while the normal pressure regulated CCN-NP exhibiting an over-high -C≡N concentration and the traditional pyrolysis CN-B showing abundant hydrogen groups and low crystallinity.

The optimized interfacial charge transport of CCN-P has also been directly evidenced by the pseudo-first-order rate constant (*k*) of surface photocarrier recombination using the following equation based on the decay profiles of the open-circuit potential: [43] [14]

$$(V - V_{light})/(V_{dark} - V_{light}) = 1 - \exp(-kt) \qquad (2)$$

Where V, $V_{dark}$, and $V_{light}$ are the open-circuit potential in dark and under irradiation, respectively (**Figure 6c**). In good accordance with $R_{trap}$, CCN-P also presents the slowest decay kinetics (0.037 s$^{-1}$) compared to CN-B (0.190 s$^{-1}$) and CCN-NP (0.044 s$^{-1}$, **Figure 6e**). Due to the above enhanced solar light (**Figure 5b-e**), lower charge transfer resistance (**Figure 6b**), and promoted photocarrier separation in both bulk phase and surface (**Figure 5a, Figure 6c**), CCN-P achieved the highest photocurrent of 573.4 A cm$^{-2}$, which is around 5.0 and 16.0 times higher than CCN-NP and CN-B, respectively (**Figure 6d**). Therefore, we draw a preliminary conclusion that our high-pressure regulated high crystalline CCN-P enables fast charge transport in both bulk phase and surface/aqueous solution interface, which accelerates the redox kinetics and maximizes the positive side of defect engineering.

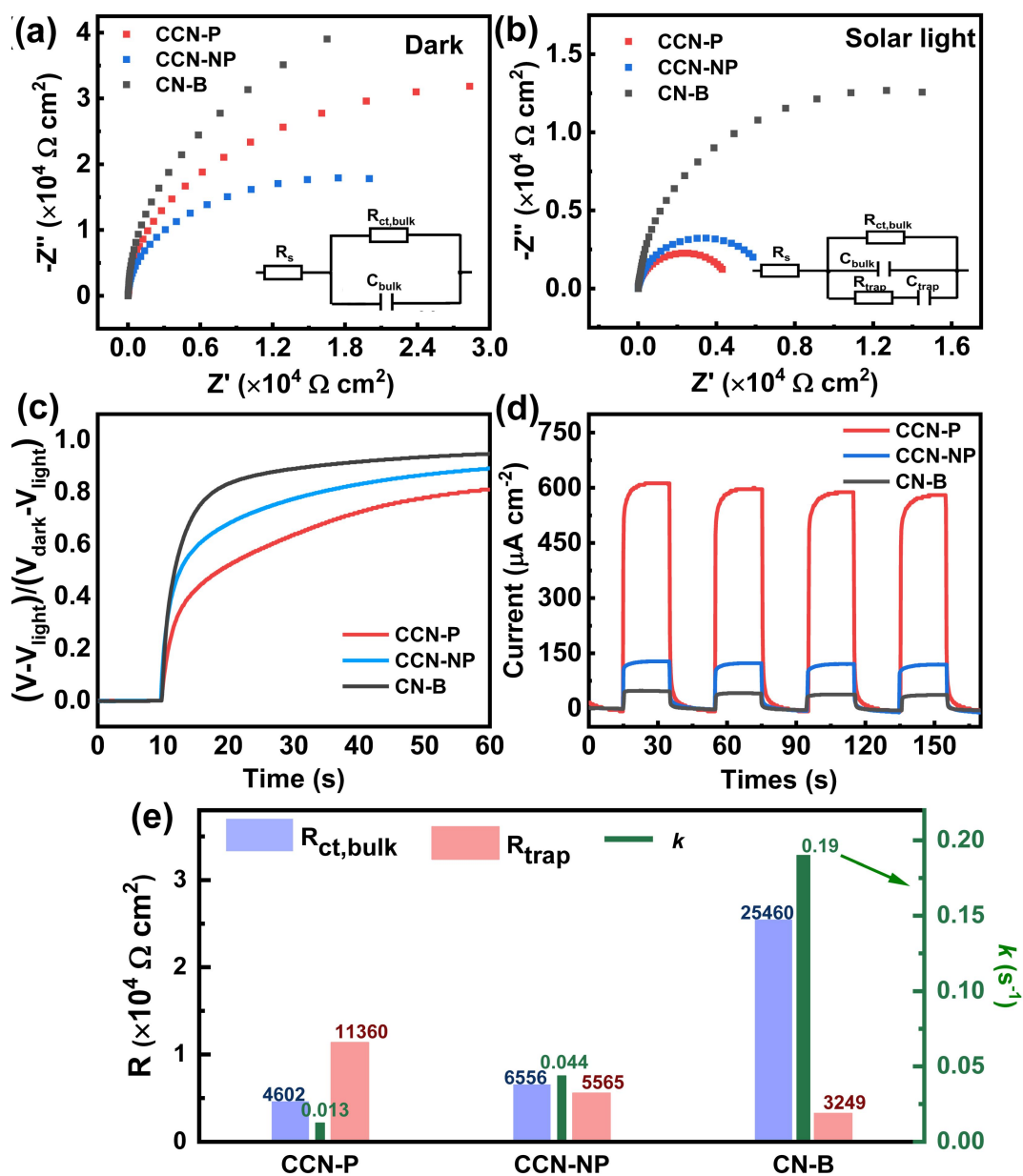

**Figure 6** EIS data of CCN-P, CCN-NP, and CN-B in **(a)** Dark; **(b)** Solar irradiation; **(c)** Open-circuit potential decay curves; **(d)** Photocurrent density and **(e)** Summary of $R_{ct,bulk}$, $R_{trap}$, and $k$ of all g-$C_3N_4$ samples.

**Photocatalytic $H_2$ evolution performance**

We first evaluated the solar application of $H_2$ evolution rate (HER) for CCN-P, CCN-NP, and CN-B using the 10% TEOA aqueous solution as reaction media. As reflected by the time course plots of $H_2$ production in **Figure 7a**, CCN-P shows the most steadily increasing trend with a rate of 2168.8 μmol g$^{-1}$ h$^{-1}$, which is 2.55 and 4.86-fold higher than those of CCN-NP and CN-B, respectively (**Figure 7b**). This is in line with PEC results, further demonstrating the critical role of high-pressure regulated ion thermal reaction in maintaining both the high crystallinity and optimized surface states of g-$C_3N_4$. Although the surface area is relatively low at only 5.81 m$^2$ g$^{-1}$, its apparent HER (ratio of HER/surface area) is unexpectedly high at 373.3 μmol m$^2$ h$^{-1}$, far more exceeding its counterparts (**Figure 7c**) and previous literature (**Figure**

**7 d)**.[26,44–52] This strongly confirms the superior HER performance of these exposed active sites in CCN-P due to the high-pressure synthetic strategy.

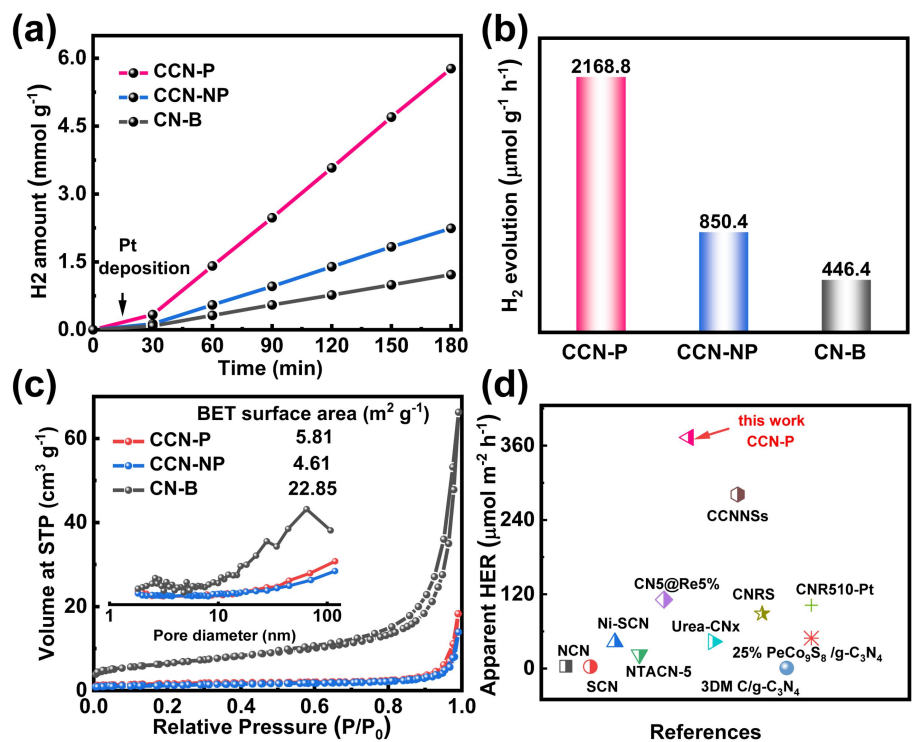

**Figure 7 (a)** Time course plots; **(b)** $H_2$ evolution column from water containing 10 vol% TEOA scavengers under simulated sunlight; **(c)** $N_2$ adsorption-desorption isotherms (inset: corresponding pore size distribution plots and BET surface area) of CCN-P, CCN-NP, and CN-B; **(d)** Comparison of apparent HER of CCN-P and previous reports. [26,44–52]

**2.5 Cycling performance of CCN-P after $H_2$ evolution**

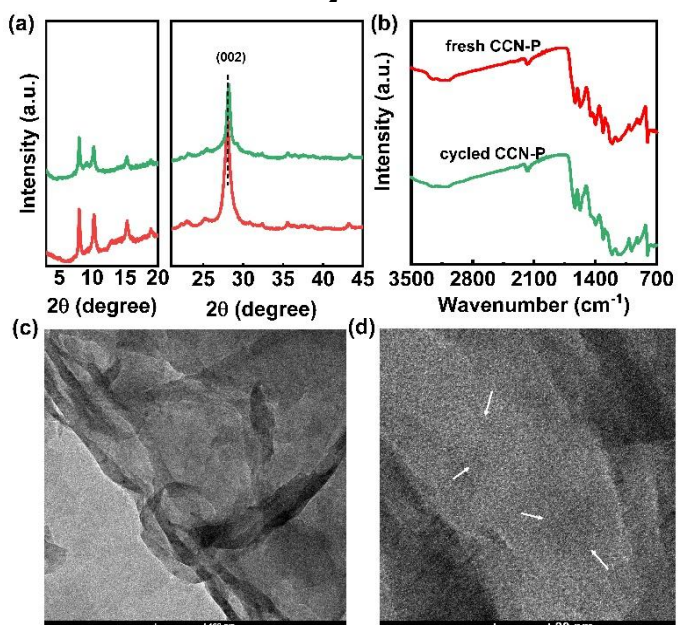

**Figure 8** Comparison of **(a)** XRD patterns and **(b)** FT-IR between fresh CCN-P and cycled CCN-P after photocatalytic $H_2$ evolution. **(c, d)** HRTEM images of cycled CCN-P.

The stability of CCN-P after photocatalytic H₂ evolution was further investigated by performing the XRD patterns, FT-IR spectra, and HRTEM images. As shown in **Figure 8a**, the obvious (100) and (002) planes of cycled CCN-P can still be clearly evidenced, indicating its excellent crystallinity stability. Similar phenomena on the functional groups and nanostructure can also be revealed in **Figure 8b-d**, further demonstrating the superior physicochemical properties.

### 2.6 Photocathodic protection of 304 ss

The photocathodic protection (PCP) of metals (such as 304 ss, carbon steel) is another important application for g-C$_3$N$_4$-based materials, and this sustainable anti-corrosion strategy is also closely associated with the solar harvesting ability and photocarrier separation efficiency of photocatalysts. We employed a three-electrode corrosion cell with the 304 ss as the working electrode in 3.5 wt% NaCl solution and a PEC cell with the as-prepared g-C$_3$N$_4$ samples as the coupled photoelectrode in 0.2 M NaOH/0.1 M Na$_2$S solution, of which the two cells are connected via a salt bridge. Once irradiation, the photoexcited electrons from CCN-P instead of the lost electrons from 304 ss would participate in the inevitable ambient oxidation, thus preventing the metal corrosion process.

It is widely accepted that the open circular potential (OCP) is the most important indicator to describe the tendency of 304 ss corrosion, and a more negative OCP value normally implies a better anti-corrosion property.[53–58] Consisting with the previous report, the OCP of 304 ss is also around -0.189 V *vs.* AgCl/Ag electrode (**Figure 9a**). In sharp contrast, under irradiation, CCN-NP shows an instant much more negative potential of -0.668 V, which is 73 and 241 mV, than CCN-NP and CN-B, respectively. This further demonstrates the superior anti-corrosion tendency of the pressure-regulated crystalline CCN-P. In addition, in increasing order, CN-B, CCN-NP, and CCN-P exhibit intermittent photocurrents of 3.26, 14.43, and 20.55 uA cm$^{-2}$, which also reveals there are more photoexcited electrons transferring from CCN-P onto 304 ss electrode (**Figure 9b**). The Tafel polarization plots of bare 304 ss and 304 ss coupled with various g-C$_3$N$_4$ photoelectrodes are displayed in **Figure 9c**. It is obvious that the corrosion potential (E$_{corr}$) of 304 ss is around -0.202 V, and a negative E$_{corr}$ shift of all these g-C$_3$N$_4$ coupled electrodes is also observed, of which CCN-P exhibits the highest value by -0.440 V. Along with the higher corrosion current (i$_{corr}$) of CCN-P (8.28 uA cm$^{-2}$) than CN-B (1.41 uA cm$^{-2}$) and CCN-NP (6.40 uA cm$^{-2}$), these result further indicates the higher photoelectronic conversion efficiency and best anti-corrosion property of CCN-P, which is also in good agreement with OCP and photocurrent tests (**Figure 9a, b**). The excellent PCP protection stability of CCN-P was also recorded by the long-term irradiated OCP values. As shown in **Figure 9d**, CN-B indicates the lowest OCP retention rate of only 57.12% with a dramatical potential drop in the first 2000 s from -0.489 V to -0.286 V. While CCN-P exhibits the highest retention rate of 78.5% with a slight OCP drop from -0.651 V to -0.519 V, exceeding by 12.0% to CCN-NP, which is supposed to be the combined effect of high crystallinity and propitiate defect control of surface groups.

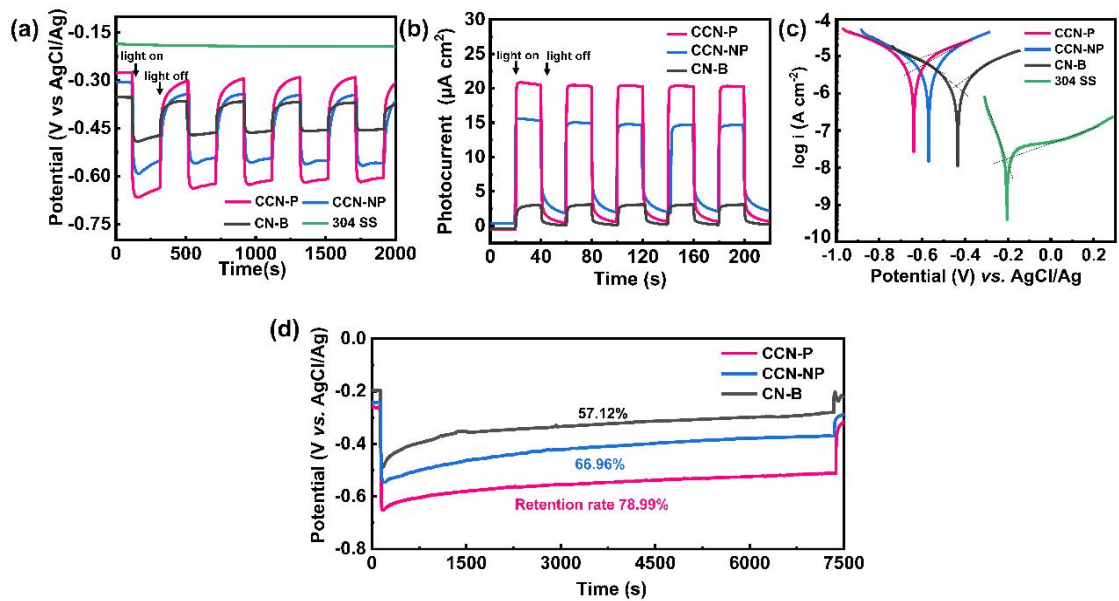

**Figure 9** **(a)** OCP curves and **(b)** Photocurrent response under intermittent solar irradiation of CN-B, CCN-P, and CCN-NP; **(c)** Tafel plots of various g-$C_3N_4$ photoelectrodes coupled to 304 ss electrode under irradiation and bare 304 ss electrode; **(d)** Long-time OCP stability test of 304 ss electrode coupled with CN-B, CCN-P, and CCN-NP.

**2.7 Cycling performance of CCN-P after 304 ss protection**

The cycling stability of CCN-P and morphology/chemical states change of 304 ss are critical for the overall evaluation of the metal anti-corrosion effect for our high-pressure and ion thermal regulated g-$C_3N_4$. To this end, the crystalline structure and morphology of cycled CCN-P were further studied by XRD and HRTEM techniques. As displayed in **Figure 10a**, the cycled CCN-P exhibits almost the same XRD patterns with the typical (100) and (002) planes at around 7.9° and 28.1°, respectively. It also shows a nanosheet structure with lattice fringe, which is consistent with that of fresh CCN-P (**Figure 10b, Figure 2e-f**). Therefore, one can conclude that CCN-P is a robust photocatalyst for both water splitting (**Figure 8**) and 304 ss protection. Besides, no obvious morphology difference is observed for 304 ss before and after the CCN-P-based photocathodic protection according to almost the same metallographic microscope images in **Figure 10c-d**, which chose the same picture capturing area. Being the most important elements in 304 ss, Fe, Cr, Ni, and O elements all present similar core-level XPS spectra before and after the PCP test (**Figure 10e-h**), further demonstrating the superior photocathodic protection of CCN-P material.

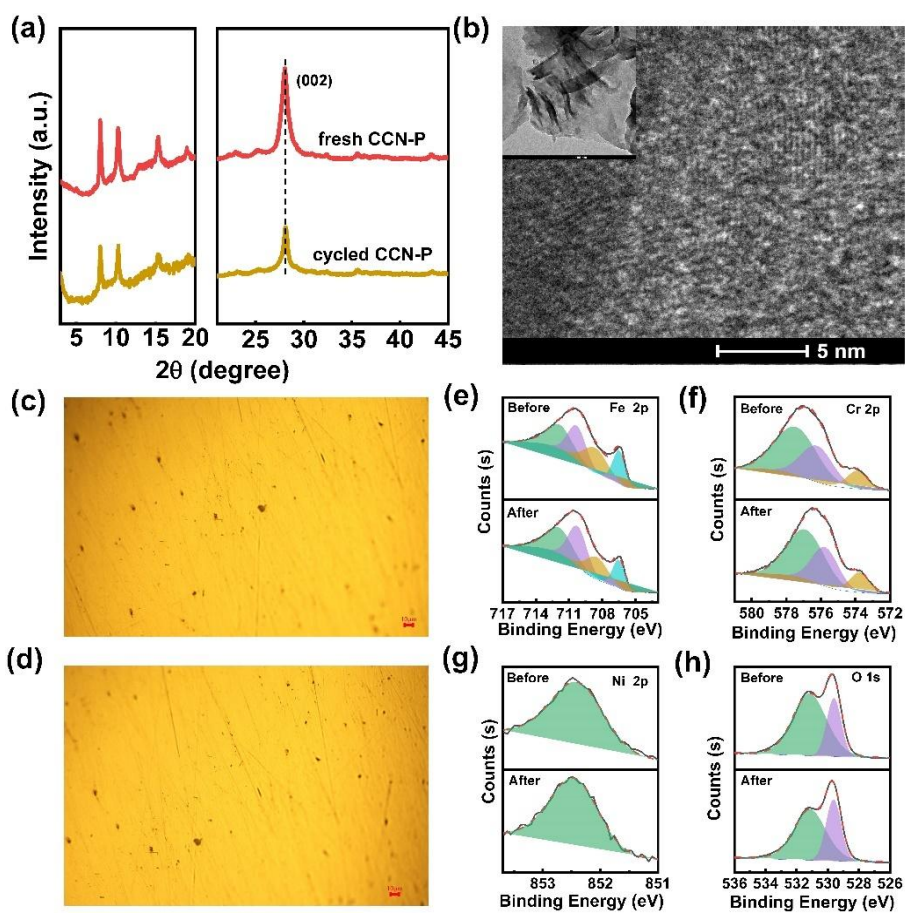

**Figure 10 (a)** Comparison of XRD patterns between fresh CCN-P and cycled CCN-P after 304 ss PCP activity; **(b)** TEM image of cycled CCN-P; Metallographic microscope images of 304 ss surface **(c)** before and **(d)** after PCP test within the same area; Core-level XPS spectra of **(e)** Fe 2p; **(f)** Cr 2p; **(g)** Ni 2p and **(h)** O 1s for 304 ss before and after PCP test.

## 3. Conclusions

In summary, we have successfully fabricated a high crystalline CCN-P with appropriate surface defects via the high-pressure regulated ion thermal strategy in the presence of NaCl/KCl eutectic salts. In comparison with the counterpart CCN-NP without pressure modification, the high strain inside the precursor tablet rendered CCN-P with a narrower interlayered distance by 0.258 nm, which is favorable to shorten the pathways of active species diffusion and accelerate the photocarrier transport kinetics from bulk to surface. Furthermore, the high pressure also imparts CCN-P with an appropriate amount of -C≡N and -NH$_x$ functional groups, which not only extend its solar light harvesting ability but also optimize its surface states with a moderate electron trapping ability. Therefore, CCN-P exhibits the highest electron-trapping resistance ($R_{trap}$) of 11.36 kΩ cm$^2$ and the slowest surface photocarrier decay kinetics of 0.013 s$^{-1}$, demonstrating a better spatial charge transfer from bulk phase to surface. Due to the high crystallinity and optimized surface states, CCN-P delivers a superior apparent photocatalytic H$_2$ evolution rate of 373.3 μmol m$^{-2}$ h$^{-1}$, ranking to the top among previous reports. In addition, it also exhibited a stable anti-corrosion property of 304 ss with the highest OCP retention rate of 78.5% after 7500 s from -0.668 V, far more exceeding the bulk g-$C_3N_4$ and CCN-P. Therefore, this high-pressure mediated ion thermal strategy provides us with a new insight for balancing both the crystallinity and spatial charge transport, which is believed to inspire more functional g-$C_3N_4$-based materials for various solar applications.

# Supporting Information

# Pressure-mediated crystalline g-$C_3N_4$ with enhanced spatial charge transport for solar $H_2$ evolution and photocathodic protection of 304 stainless steels

1. **Experimental Section**

**1.1 Materials**

Melamine ($C_3H_6N_6$, 99%), NaCl ( ACS reagent, ≥99.0%), KCl (ACS reagent, ≥99.0%), and Nafion$^{TM}$ 117 solution (~5% in a mixture of lower aliphatic alcohols and water) were purchased from Sigma-Aldrich. All chemicals were used as received without further purification.

**1.2 Synthesis of CCN-P and CCN-NP**

The CCN-P was synthesized via the ion thermal pyrolysis of melamine, NaCl, and KCl in the form of a tablet. Typically, a mixture of NaCl and KCl was well ground in a mortar at a mole ratio of 76:24, followed by a heating treatment under 200°C. After cooling to room temperature, melamine and NaCl/KCl were mixed in a mass ratio of 1:4. Then, the white powder was pressed into a tablet with a diameter of 12 mm under a pressure of 7 tons for 5 min. The tablets were wrapped with aluminium foil, transferred into a porcelain combustion boat, and then heated at 550 °C for 4 h at a ramping rate of 5 °C min$^{-1}$ in Ar atmosphere. After cooling, the mixture was finely ground and dissolved in hot water. Finally, the light-yellow or yellowish-green powder was obtained after filtering, washing, and drying. The resulting crystalline g-$C_3N_4$ with high-pressure regulation was named CCN-P. The controlled sample of crystalline g-$C_3N_4$ without pressure regulation (CCN-NP) was pyrolyzed according to the same procedures except for the precursor in the form of loosened powder.

**1.3 Synthesis of bulk g-$C_3N_4$ (CN-B)**

To make a good comparison, CN-B was obtained by calcining melamine at 550 °C for 4 h at a ramping rate of 5 °C min$^{-1}$ in the Ar atmosphere.

**1.4 Preparation of photoelectrodes**

For the synthesis of photoelectrodes, 10 mg of the as-prepared materials were dissolved in 1 mL of ethanol: Nafion (9: 1, by volume) and sonicated for 10 min. Then 30 uL of the suspension was coated on a 10 mm × 15 mm fluorine-doped tin oxide (FTO) conducting glass by a spin coater (Laurell WS-400A-6NPP Lite, USA) within 2 min at 4000 r.p.m. The electrodes were then transferred into a tubular furnace, followed by thermal treatment at 200 °C for 4 h under an $N_2$ atmosphere.

**1.5 Structural Characterizations**

The scanning electron microscopy (SEM) images were collected on a Zeiss Supra 55VP microscope operating under an acceleration voltage of 5~15 kV. The high-resolution transmission scanning electron microscopy (HRTEM) images were observed using a Tecnai-G2 F30 S-Twin microscope equipped with an energy-dispersive X-ray spectroscopy spectrometer. The X-ray diffraction (XRD) patterns were analyzed by a Bruker D8 diffractometer with Cu Kα radiation ($\lambda$ = 1.5418 Å). The Fourier transform infrared (FT-IR) spectra were acquired by a Nicolet 6700 instrument. The $N_2$ adsorption and desorption isotherms were carried out employing a Micromeritics 3 Flex analyzer. The X-ray photoelectron spectroscopy (XPS) measurements were studied by an ESCALAB250Xi spectrometer with an achromatic Al Kα source, and all peaks were regulated at 284.6 eV. The ultraviolet-visible diffuse reflectance spectra (UV-vis DRS) were investigated by a Perkin Elmer Lambda 950 spectrophotometer. The electron paramagnetic resonance (EPR) was performed on a Bruker A300-10/12 EPR spectrometer with f=9.853 GHz. The photoluminescence (PL) spectra were carried out using a Varian Eclipse Fluorescence Spectrometer.

**1.6 Photoelectrochemical measurements**

The photoelectrochemical performance was measured in a standard three-electrode cell on a CHI 650D electrochemical workstation (CHI Instrument, USA). The FTO photoelectrode, platinum wire, and Ag/AgCl (saturated KCl) electrode were employed as the working electrode, counter electrode, and reference electrode in an electrolyte of 0.35 M $Na_2S$/0.25M $Na_2SO_3$ aqueous solution, respectively. A Xe arc lamp (PLS-SXE300C) equipped with an AM 1.5 G filter or UV-cutoff filter ($\lambda$ > 420 nm) was employed as the light source. The light was illuminated on the backside of the FTO electrode with an intensity of 100 mW cm$^{-2}$. Photovoltammogram curves and photocurrents (i-t) were measured in a potential range from -0.4 V to 0.8 V at a scan rate of 10 mV s$^{-1}$. Electrochemical impedance spectroscopy (EIS) was carried out by applying a 5 mV alternative signal with a frequency of 100 kHz~0.01 Hz. The Mott-Schottky plots were analyzed in 0.5 M $Na_2SO_4$ aqueous solution (pH=7) and a frequency of 1~3 kHz with an alternating voltage magnitude of 10 mV. The calculated flat potential $E_f$ *vs.* AgCl/Ag can be converted to $E'_f$ *vs.* RHE via the following equation (1):[1]

$$E'_f = E_f + 0.197 + 0.059 pH \qquad (1)$$

**1.7 Photocatalytic hydrogen evolution**

The photocatalytic performance of the synthetic catalysts was assessed by the hydrogen evolution rate at room temperature. Typically, 80 mg of the samples were firstly dispersed in 80 mL of triethanolamine and water solution (volume ratio of 1:9) under sonication. Afterwards, a certain amount of $H_2PtCl_6$ (1 wt% Pt) was added to the solution to act as the co-catalyst. Then the solution was aerated with nitrogen for 0.5 h to eliminate the air, and the flask was sealed with silicone rubber and parafilm. The light source is a Xe arc lamp (PLS-SXE300C) equipped with an AM 1.5 G filter or UV-cutoff filter ($\lambda$ > 420 nm) to trigger the photocatalytic reaction. The mixture was under continuous magnetic stirring, and the flask was put in front of the lamp, where the light intensity was measured at ~ 100 mW cm$^{-2}$. The amount of evolved $H_2$ was analyzed by gas chromatography (GC2014, Shimadzu) after intermittently taking out 0.2 mL of gas.

**1.8 Photocathodic protection of 304 ss**

The photocathodic protection of 304 ss was also evaluated in a three-electrode corrosion cell coupled with a photoelectrochemical cell. The corrosion cell employed the 304 ss electrode with a diameter of 10 mm as the working electrode, the Ag/AgCl (saturated KCl) electrode as the reference electrode, and Pt wire as the counter electrode, respectively. The electrolyte is 3.5wt% NaCl solution. While in the photoelectrochemical cell, the various g-$C_3N_4$ photoelectrodes were the only electrode coupled with 304 ss via a salt bridge using the 0.1 M $Na_2S$/0.2 M NaOH as electrolyte. The open circle potential, photocurrent, and Tafel polarization tests were also performed with the above set-up under solar irradiation. The polarization test was analyzed in a potential range of -0.3V~0.3V *vs.* OCP at a scan rate of 0.5 mV $s^{-1}$.

## 2. Tables

**Table S1** Surface compositions of g-$C_3N_4$ samples obtained from XPS analysis (at. %).

| Sample | $C^1$ | $N^1$ | $O^1$ | $Na^1$ | $K^1$ | $Cl^1$ | $C/N^2$ |
|---|---|---|---|---|---|---|---|
| **CCN-P** | 38.59 | 52.84 | 4.57 | 1.78 | 1.66 | 0.56 | 0.73 |
| **CCN-NP** | 38.61 | 54.32 | 3.48 | 1.77 | 1.47 | 0.54 | 0.71 |
| **CN-B** | 42.03 | 56.52 | 1.45 | -- | -- | -- | 0.74 |

[1]Note: 1 is the atomic content (%); 2 is the atomic ratio of C/N.

**Table S2** The fractions of various C, and N species of CCN-P, CCN-NP, and CN-B (at. %).

| Species<br>Sample | C1 | C2 | C3 | N1 | N2 | N3 | N4 |
|---|---|---|---|---|---|---|---|
| **CCN-P** | 19.7 | 8.9 | 71.4 | 72.6 | 12.5 | 8.6 | 6.3 |
| **CCN-NP** | 15.3 | 11.8 | 72.9 | 74.4 | 12.3 | 9.4 | 3.9 |
| **CN-P** | 12.8 | -- | 87.2 | 57.1 | 24.8 | 14.3 | 3.8 |

**Table S3** Fitting parameters of g-C₃N₄ samples from time-resolved fluorescence spectra.

| Sample | $\tau_1$ | $A_1$ | Rel % | $\tau_2$ | $A_2$ | Rel % | $\tau_3$ | $A_3$ | Rel % | $\tau_{ave}$ |
|---|---|---|---|---|---|---|---|---|---|---|
| CN-B | 1.948 | 0.533 | 55.4 | 7.700 | 0.082 | 33.6 | 48.680 | 0.004 | 11.3 | 8.78 |
| CCN-P | 0.208 | 0.331 | 43.2 | 0.909 | 0.070 | 39.8 | 3.825 | 0.007 | 17.0 | 1.10 |
| CCN-NP | 0.035 | 1.007 | 28.9 | 0.426 | 0.418 | 47.9 | 2.454 | 0.012 | 23.2 | 0.61 |

**Table S4** EIS results of various g-C₃N₄ samples measured in dark.

| Samples | $R_s$ (Ω cm²) | $R_{ct,bulk}$ (Ω cm²) | $C_{bulk}$ (F cm⁻²) |
|---|---|---|---|
| CCN-P | 9.14 | 6.27×10⁴ | 2.62×10⁻⁵ |
| CCN-NP | 8.71 | 3.56×10⁴ | 2.54×10⁻⁵ |
| CCN-B | 13.53 | 1.07×10⁵ | 2.22×10⁻⁵ |

Note: The fitting parameters are based on the following equivalent circuit, which $R_s$ and $R_{ct,\,bulk}$ represent the solution (electrolyte) resistance and bulk charge transfer resistance, respectively. While $R_{trap}$ is trapping resistance, another critical descriptor is to evaluate the recombination of photocarriers on the surface states induced by the defects. The $C_{trap}$ is the capacitance associated with trapped photocarriers at the surface state.

**Table S5** EIS results of various g-C$_3$N$_4$ samples measured under solar irradiation.

| Samples | R$_s$ (Ω cm$^2$) | R$_{ct,bulk}$ (Ω cm$^2$) | R$_{trap}$ (Ω cm$^2$) | C$_{bulk}$ (F cm$^{-2}$) | C$_{trap}$ (F cm$^{-2}$) |
|---|---|---|---|---|---|
| **CCN-P** | 9.17 | 4.60×10$^3$ | 1.14×10$^4$ | 3.23×10$^{-5}$ | 4.83×10$^{-6}$ |
| **CCN-NP** | 8.71 | 6.56×10$^3$ | 5.57×10$^3$ | 4.57×10$^{-5}$ | 5.03×10$^{-6}$ |
| **CN-B** | 13.54 | 2.55×10$^4$ | 3.25×10$^3$ | 3.12×10$^{-5}$ | 1.00×10$^{-5}$ |